\documentclass[showpacs,aps,twocolumn]{revtex4}
\begin{document}
\title{Observation of Single-Photon Switching}
\author{Yong-Fan Chen, Zen-Hsiang Tsai, Yu-Chen Liu, and Ite A. Yu}
\email{yu@phys.nthu.edu.tw}
\affiliation{Department of Physics, National Tsing Hua University, Hsinchu, Taiwan 300, Republic of China}
\begin{abstract}
We report an experimental demonstration of single-photon switching in laser-cooled $^{87}$Rb atoms. A resonant probe pulse with an energy per unit area of one photon per $\lambda^2/2\pi$ propagates through the optically thick atoms. Its energy transmittance is greater than 63\% or loss is less than $e^{-1}$ due to the effect of electromagnetically induced transparency. In the presence of a switching pulse with an energy per unit area of 1.4 photons per $\lambda^2/2\pi$, the energy transmittance of the same probe pulse becomes less than 37\% or $e^{-1}$. This substantial reduction of the probe transmittance caused by single switching photons has potential applications in single-photon-level nonlinear optics and the manipulation of quantum information.
\end{abstract}
\maketitle 
Compared to atoms and molecules, photons are superior carriers of quantum information because they are more inert with respect to the environment. Photons can be used to transport information from quantum logic gates to quantum memory \cite{LtQt1,LtQt2,LukinEit}. The recent development of slow light arising from quantum interference has made storage and retrieval of quantum information by photon pulses feasible \cite{StopLtTh,HauStopLt,WalsworthStopLt,ScullyStopLtTh,LukinStorageLt} and low-light-level nonlinear optics practical \cite{LowLtNLO,HarrisPsExp,ZhuXpm,Zhu6Waves,Harris4Waves}. Control of one single-photon pulse by another provides more degrees of freedom in the manipulation of quantum information. The interaction between two single-photon pulses can potentially create entangled photon pairs and quantum phase gates for applications in quantum teleportation \cite{QtTh,QtExp,QtNonlinearKerr} and quantum computation \cite{XpmTh,QcRv}. Harris and Yamamoto \cite{PsTh} have suggested the system of photon switching by quantum interference in which switching of a single-photon pulse by another can be realized and single-photon-level nonlinear optics is achievable.

In the four-level system of photon switching by quantum interference, a weak probe field and a strong coupling field form the three-level $\Lambda$-type configuration that gives rise to electromagnetically induced transparency (EIT) \cite{EitExp,CptRv,HarrisEitRv}. Absorption of the probe field is suppressed or completely inhibited by the EIT effect. A switching field drives the transition between the ground state of the coupling field and another excited state. The presence of this field enables the absorption of the probe field and induces the three-photon transition from the ground level of the probe field to the excited level of the switching field. With the four-level system, Harris and Yamamoto obtained the important result that one switching event of the probe field costs one switching photon under ideal conditions. This result leads to the intriguing concept of the threefold entangled state depicted in Fig.~3 of Ref.~\cite{PsTh}. The phenomenon of photon switching by quantum interference in a four-level atomic system has been observed by Yan {\em et al}. \cite{ZhuPsPra} and Braje {\em et al}. \cite{HarrisPsExp}, and its transient behavior in a simple four-state system has been studied by Chen {\em et al}. \cite{TransientPs}. However, single-photon switching was not achieved in those experiments. This paper reports the first observation of photon switching at the single-photon level.

We studied photon switching in cold $^{87}$Rb atoms produced by a vapor-cell magneto-optical trap (MOT). 
Typically, we trapped $1.2\times10^9$ in our MOT, as measured by the optical-pumping method \cite{AtomNumber}. The size of the atom cloud was approximately $3.2\times3.2\times 2.6$ mm$^3$.
The coupling, switching, and probe fields came from three diode lasers, all of which were injection-locked by the same master laser. 
One beam from the master laser was sent through a 6.8-GHz electro-optic modulator (EOM) (New Focus 4853). The high-frequency sideband of the EOM seeded the probe laser. The probe beam had a $1/e$ diameter of 5.5 mm and passed through a 1-mm aperture before interacting with atoms. The intensity or the Rabi frequency of the probe field was determined from the ratio of the power to the area of the aperture. The probe, coupling, and switching fields resonantly drove $|F=1\rangle \rightarrow |F'=2\rangle$, $|F=2\rangle \rightarrow |F'=2\rangle$, and $|F=2\rangle \rightarrow |F'=3\rangle$, respectively.
($F$ and $F'$ denote hyperfine levels in the $5S_{1/2}$ ground state and $5P_{3/2}$ excited state.)
Each of them passed through a distinct acousto-optic modulator (AOM) that could be switched on or off individually. The driving frequencies of these three AOMs were maintained constant throughout the experiment. The coupling and probe fields were circularly polarized with right helicity ($\sigma_+$ polarization). The switching field was also circularly polarized, but with left helicity ($\sigma_-$ polarization). 
Considering the Zeeman sublevels, our system consisted of three sets of four-state subsystems.
All three fields propagated in nearly the same direction.

The timing sequence of the measurement is described below. We first turned off the magnetic field of the MOT and then, after a 1.45-ms delay, shut off the repumping beam and switch on the coupling field. The trapping beams were turned off 45 $\mu$s later to prevent the population from being trapped in the $|F=2,m=2\rangle$ state. The entire population was optically pumped to the $|F=1\rangle$ ground state. Finally, the probe and switching fields were switched on. A 125-MHz photodiode (New Focus 1801) detected the probe transmission and its output was directly sent to a digital oscilloscope. After the measurement was complete, the probe, coupling, and switching fields were turned off and the MOT was turned back on. The above sequence was repeated with a period of 100 ms; data were averaged 128 times by the oscilloscope before being transferred to the computer.

We measured the steady-state EIT spectrum of the probe transmission to determine the Rabi frequency of the coupling field, $\Omega_c$. The atom number was reduced in this measurement to identify the absorption maxima. 
Figure~\ref{Spectra} shows a typical EIT spectrum and the probe transmission in the absence of the atoms. 
We scanned the probe frequency linearly at a rate of 2.5 $\Gamma$/ms by ramping the EOM driving frequency. The sweep rate was slow enough that the narrow transparency window could be clearly resolved. Since the atom cloud gradually expands in the absence of the trap or the cold atoms get pushed away by the laser fields during the spectroscopic measurement, the spectrum is slightly asymmetric \cite{FastEit}. From the separation of the two absorption maxima in the spectrum, we obtained $\Omega_c = 0.40\Gamma$. This value is about 8\% higher than the average of the estimates of the Rabi frequencies of the three subsystems based on the laser power and beam profile. All of the $\Omega_c$ values quoted in this paper are determined with the spectroscopic measurement. Since the average Rabi frequency of the probe field, $\Omega_p$, is $0.07\Gamma$, it can be considered as a weak perturbation. By comparing the measured spectrum with the theoretical prediction of a simple three-level system, we find that the relaxation rate $\gamma$ of the ground-state coherence is around $0.002\Gamma$ in our system. This value is consistent with the value we obtained previously in experiments on a simple three-state system \cite{TransientPs}. The black solid line in Fig.~\ref{Spectra} shows the probe transmission under the same experimental conditions except that the switching field is present. This spectrum also reveals that the optical-pumping effect is changed by the presence of the switching field.

The optical depth of the atoms and the Rabi frequency of the switching field, $\Omega_s$, were directly determined from the delay time of the probe pulse, $T_D$. We generated the Gaussian pulse of the probe field by controlling the RF power of the AOM that switches the field. This control was achieved with a variable-gain amplifier (Mini-Circuits ZFL-1000GH) and an arbitrary waveform generator (SRS DS345). However, the Gaussian pulse width was limited to about 1 $\mu$s. Throughout the experiment, the energy per unit area of the probe pulse was always about one photon per $\lambda^2/2\pi$. We set the peak of the input probe pulse $\Omega_{p,max}$ to $0.10\Gamma$ and the pulse width to 2.4 $\mu$s. 
Figure~\ref{DelayTime} shows the experimental data along with the best fits of the Gaussian function. 
For the condition that all the fields are on the resonance and $\gamma$ is negligible, the delay time is given by \cite{PsTh}
\begin{equation}
	T_D = N\sigma L \frac{\Gamma\Omega_c^2}{(\Omega_c^2+\Omega_s^2)^2},
\end{equation}
where $T_D$ is the delay time and $N\sigma L$ is the optical density of the resonant one-photon probe transition. The probe width is large enough that the steady-state solution is valid. In the measurement, $\Omega_c = 0.40\Gamma$. From the delay time in the EIT condition, we estimate that $ N\sigma L = 7.1$ in our system. From the ratio of the delay times in the EIT and switching conditions, we estimate $\Omega_s = 0.21\Gamma$. This value is about 5\% larger than the average of the estimates of the Rabi frequencies of the three subsystems based on the laser power and beam profile. All the $\Omega_s$ values quoted in this paper are determined with the delay-time measurement. Unless specified, the energy per unit area of the switching pulses is approximately one photon per $\lambda^2/2\pi$ or $\Omega_s = 0.12\Gamma$ of a 2.4-$\mu$s square pulse throughout the entire experiment. 

Figure~\ref{Efficiency}(a) plots the switching efficiency for different pulses of the switching field.
We set $\Omega_c = 0.69\Gamma$ and the probe pulse width to 4.8 $\mu$s. The switching efficiency is optimized when the width of the square switching pulse matches the $1/e$ width of the probe pulse. This is expected, since this combination of pulses concentrates all the energy of the switching field on the high-density part of the probe pulse. The transmitted probe pulse in the optimized switching condition shows little deviation from the Gaussian function. When the width of the switching pulse is less than or equal to half of the probe pulse width, the transmitted probe pulse is significantly distorted. Thus, for the remainder of this report we will focus on the square switching pulse with a width equal to the probe pulse width.

Figure~\ref{Efficiency}(b) shows the ratio of the EIT to switching optical densities for different probe pulse widths. In this study, the coupling intensity was proportional to the reciprocal of the probe pulse width in order to maintain good EIT bandwidth with respect to frequency bandwidth of the probe pulse as well as to keep $\Omega_{p,max}/\Omega_c$ constant and small. Since the square switching pulse width is set to the $1/e$ width of the probe pulse and its energy per unit area is constant, the intensity of the switching pulse is also proportional to the reciprocal of the probe pulse width. 
The optical density, $\alpha$, in the steady-state EIT or switching condition is given by \cite{PsTh}
\begin{equation} \label{OD}
	\alpha = N\sigma L \frac{\Omega_s^2+2\gamma\Gamma}
		{\Omega_c^2+\Omega_s^2+2\gamma\Gamma}.
\end{equation}
In the above equation, the probe frequency is assumed on the resonance. For $\Omega_c^2$, $\Omega_s^2$ $\gg$ $2\gamma\Gamma$, this equation explains that the data points of the three shorter pulses depend on $\tau$ linearly. In general, a shorter pulse width results in a better switching efficiency. However, the Gaussian shape is not ideal for pulses of widths less than 2 $\mu$s in the experiment. We chose a pulse width of 2.4 $\mu$s for the measurements shown in Figs.~\ref{DelayTime}, \ref{Efficiency}(c), and \ref{Efficiency}(d).

Figure~\ref{Efficiency}(c) plots the energy transmittance versus the number of switching photons per $\lambda^2/2\pi$. The switching pulse center is delayed 1.2 $\mu$s with respect to the probe pulse center. We fit the data with the function: 
\begin{equation} \label{fit}
	y(x) = p {\rm e}^{-\alpha(x=0)} + (1-p) {\rm e}^{-\alpha(x)},
\end{equation}
where $y(x)$ is the transmittance, $x$ is $\Omega_s^2$ in the form of energy per unit area, $\alpha(x)$ is the optical density described by Eq.~(\ref{OD}), and $p$ is the percentage of the probe energy without the interaction with the switching field. In the function of $\alpha(x)$, $\Omega_c = 0.40\Gamma$ or $\Omega_c^2 = 11.1$ in the same units of $x$, and $N\sigma L$ and $\gamma$ are the fitting parameters. For the best fit, $p = 0.15\pm0.02$, $ N\sigma L = 7.4\pm0.5$, and $\gamma = (4.2\pm0.4)\times10^{-3}\Gamma$. These values are reasonable for the experimental conditions. Figure~\ref{Efficiency}(d) shows the variation in switching efficiency as a function of the offset between the centers of the probe and switching pulses, $t_d$. Interpolation of the data indicates that the best $t_d$ should be a little larger than 0.6 $\mu$s. An offset of this magnitude is in line with expectations for a 2.4-$\mu$s square pulse to switch most of the probe energy in the medium, since the delay time of the probe pulse is 1.2 $\mu$s in the EIT condition and becomes shorter in the presence of the switching pulse.

The inset of Fig.~\ref{DelayTime} shows a demonstration of the single-photon switching. 
The square switching pulse has an energy per unit area of 1.4 photons per $\lambda^2/2\pi$ or $\Omega_s = 0.14\Gamma$ at a width of 2.4 $\mu$s. 
We set $t_d = 0.6$ $\mu$s, $\Omega_{p,max} = 0.10\Gamma$, and $\Omega_c = 0.40\Gamma$. The energy transmittance is 64.1\% in the EIT condition, but becomes 36.9\% in the switching condition. The delay times are 1.2 and 0.95 $\mu$s in the absence and presence of the switching pulse, respectively. The probe pulse in the inset shows discontinuities in its slope at two places due to the rising and falling edges of the square switching pulse.

The experimental system can be improved in the following ways. (i) Incorporation of a compressed MOT \cite{CMOT} into the measurement sequence can increase the density of the cold atoms, and hence the optical density of the one-photon probe transition. (ii) The spectral width of the master laser that injection-locks the probe, coupling, and switching lasers is about $0.5\Gamma$, including the modulation amplitude and residual fluctuation in the frequency stabilization. This laser linewidth does not influence the EIT effect, but decreases the absorption in the photon switching. Reduction of the laser linewidth may enhance the switching efficiency. (iii) The entire population can be optically pumped to the $|F=1,m=0\rangle$ state during the measurement of the probe transmittance. The experiment will then become a simple four-level system consisting of the $|F=1,m=0\rangle$, $|F'=2,m=1\rangle$, $|F=2,m=0\rangle$, and $|F'=3,m=-1\rangle$ states. 
Since the $|F=1,m=0\rangle$ and $|F=2,m=0\rangle$ states are immune to the axial magnetic field, the relaxation rate of the ground-state coherence can be further reduced by adding a stable magnetic field in the propagation direction of the three laser fields. (iv) There are AOM drivers with a higher bandwidth of RF power control. Utilizing such driver to generate shorter Gaussian pulses can improve the switching efficiency as demonstrated in Fig.~\ref{Efficiency}(b). (v) The probe and switching pulses should be coupled into the same optical fiber in order to achieve a perfect spatial overlap. After interacting with the atoms, the outgoing pulses can be distinguished by using a quarter-wave plate and a polarizing beam splitter.

In conclusion, we have systematically studied the photon switching by quantum interference in laser-cooled $^{87}$Rb atoms and demonstrated single-photon switching. The experimental system is optically thick such that the probe pulse of one photon per $\lambda^2/2\pi$ is absorbed to $e^{-7}$ by propagating through the atoms. The presence of the coupling field at a moderate intensity or $\Omega_c = 0.4\Gamma$ suppresses the one-photon absorption and reduces the energy loss of the probe pulse to less than $e^{-1}$. When the probe pulse co-propagates with the switching pulse of 1.4 photons per $\lambda^2/2\pi$ in the constant presence of the coupling field, two-photon absorption decreases the energy transmittance of the probe pulse to $e^{-1}$. Studies similar to those in Refs.~\cite{ZhuXpm,Zhu6Waves,Harris4Waves} can now be performed at the intriguing energy level of single photons. Our results show that applications exploiting single-photon-level nonlinear optics and the three-fold entangled state to manipulate quantum information are coming to realization. 

We thank Prof. Y. M. Kao of National Chiao Tung University and Prof. H. S. Chou of National Taiwan Ocean University for helpful discussions. This work was supported by the National Science Council under NSC Grant No. 93-2112-M-007-013. 

\newpage
\begin{figure}[!]
\caption{Probe transmission versus detuning in the absence of the atoms (gray solid line), and in the EIT (gray dashed line) and switching (black solid line) conditions. $\Omega_p = 0.07\Gamma$, $\Omega_c = 0.40\Gamma$, and $\Omega_s = 0.30\Gamma$.}
\label{Spectra}
\end{figure}

\begin{figure}[!]
\caption{(Color online) Probe transmissions in the absence of the atoms (black line or the largest peak amplitude), in the EIT (red line) condition, and under the constant presence of both the coupling and switching fields (blue line or the smallest peak amplitude). Solid lines represent the experimental data. Dashed lines are the best fit of the Gaussian function. $\Omega_{p,max} = 0.10\Gamma$, $\Omega_c = 0.40\Gamma$, and $\Omega_s = 0.21\Gamma$. Delay times with and without the switching field are 0.73 and 1.2 $\mu$s, respectively. The inset shows the probe transmission (blue solid line) in the presence of the square switching pulse (black dotted line). 
}
\label{DelayTime}
\end{figure}

\begin{figure}[!]
\caption{(a) Energy transmittance of the probe pulse under different switching pulses. $\Omega_{p,max} = 0.07\Gamma$ and $\Omega_c = 0.69\Gamma$. Squares and triangles represent data of the square and Gaussian switching pulses, respectively. From short to long square pulses, $\Omega_s = 0.12$, $0.085$, $0.069$, and $0.060\Gamma$. The Gaussian pulse has a peak $\Omega_{s,max}$ of $0.090\Gamma$ and a $1/e$ width of 4.8 $\mu$s. (b) Ratio of the EIT to switching optical densities versus the pulse width. $\Omega_{p,max} = 0.155\Gamma/\sqrt{\tau}$, $\Omega_c = 0.62\Gamma/\sqrt{\tau}$, and $\Omega_s = 0.22\Gamma/\sqrt{\tau} $, where $\tau$ is the probe pulse width in units of $\mu$s. (c) Energy transmittance versus number of switching photons per $\lambda^2/2\pi$. $\Omega_{p,max} = 0.10\Gamma$ and $\Omega_c = 0.40\Gamma$. Error bars are about the size of data points. Solid line is the best fit with the function in Eq.~(\ref{fit}). (d) Energy transmittance versus offset between the centers of the probe and switching pulses. $\Omega_{p,max} = 0.10\Gamma$, $\Omega_c = 0.40\Gamma$, and $\Omega_s = 0.13\Gamma$.}
\label{Efficiency}
\end{figure}
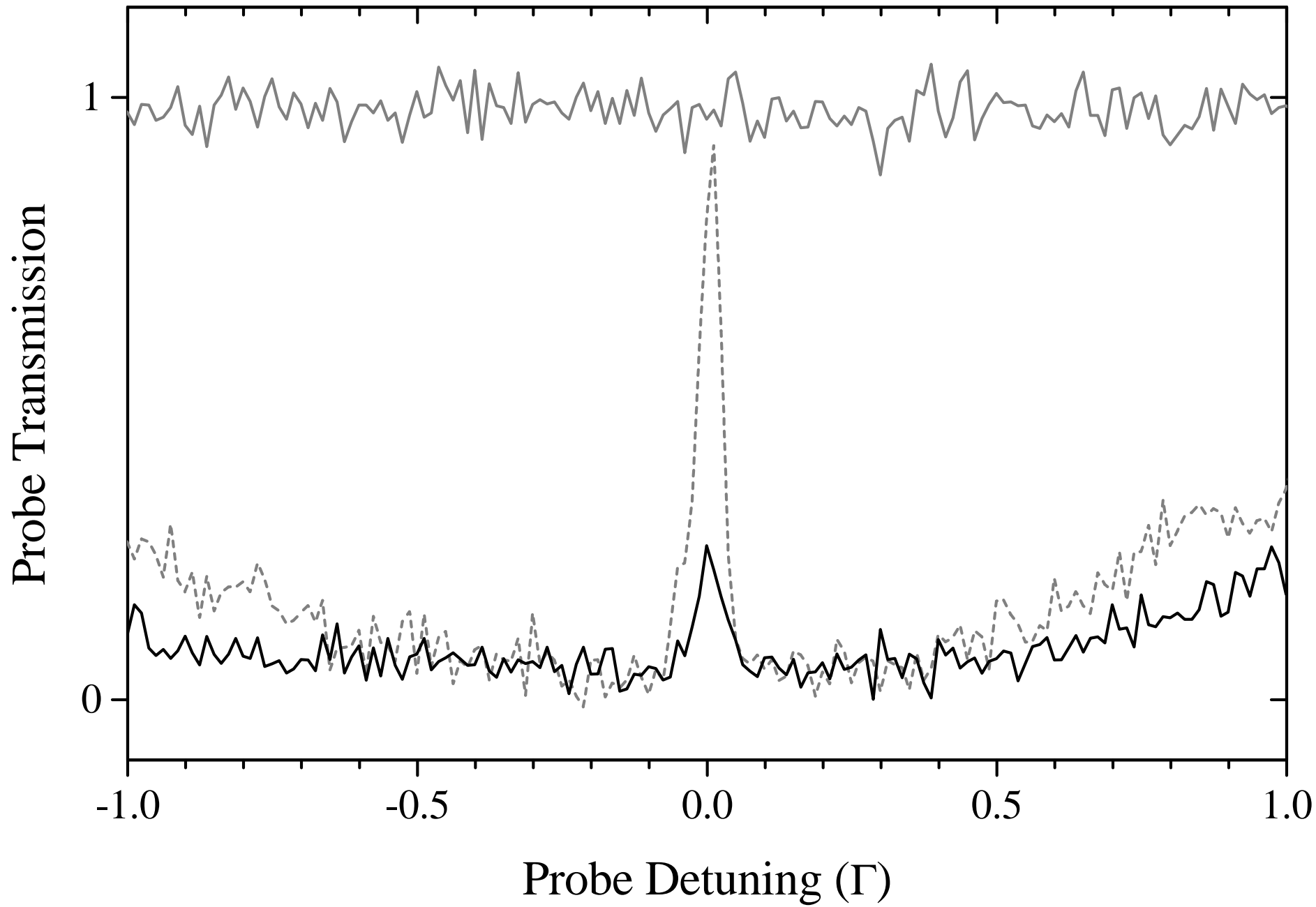
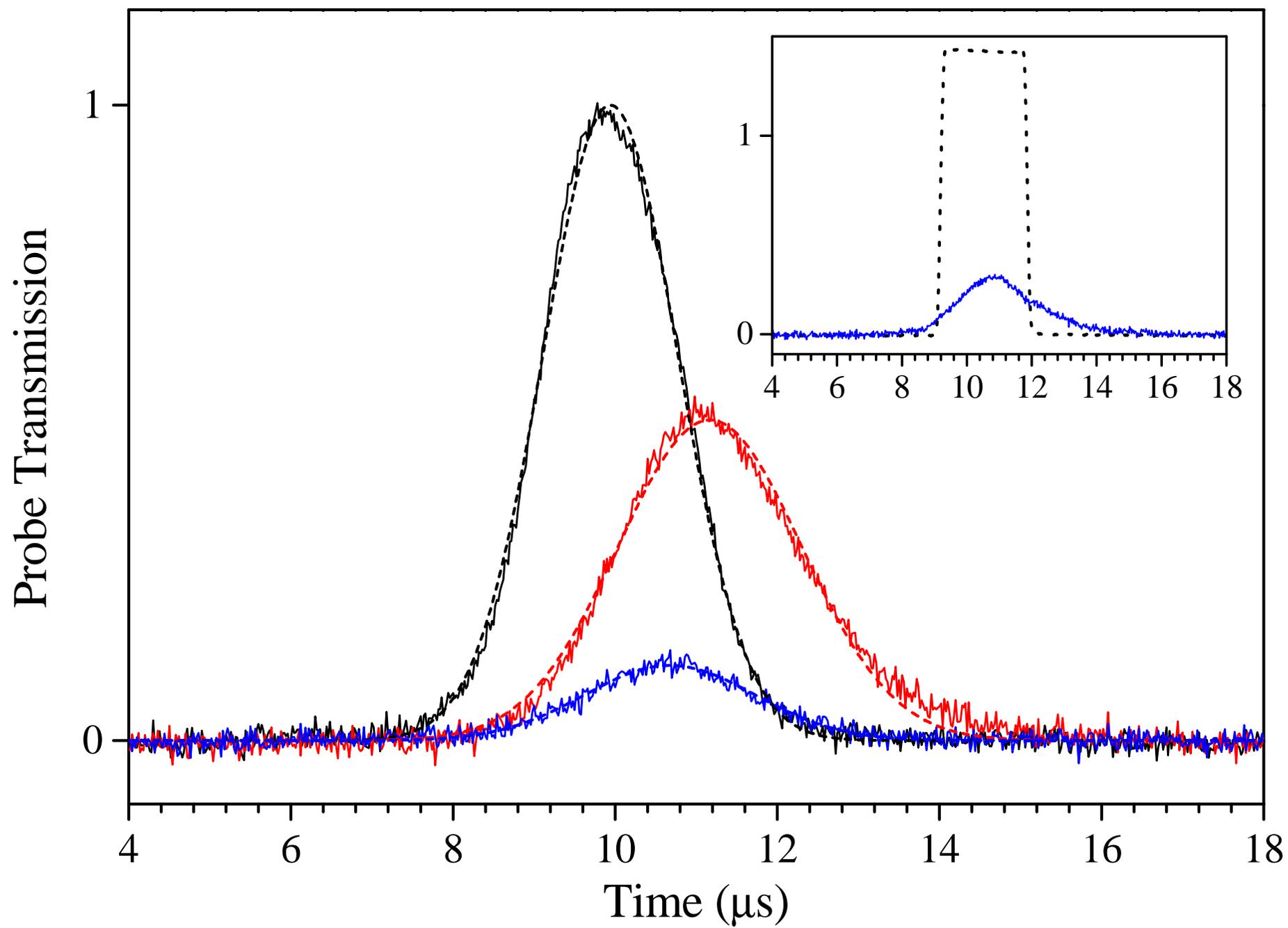
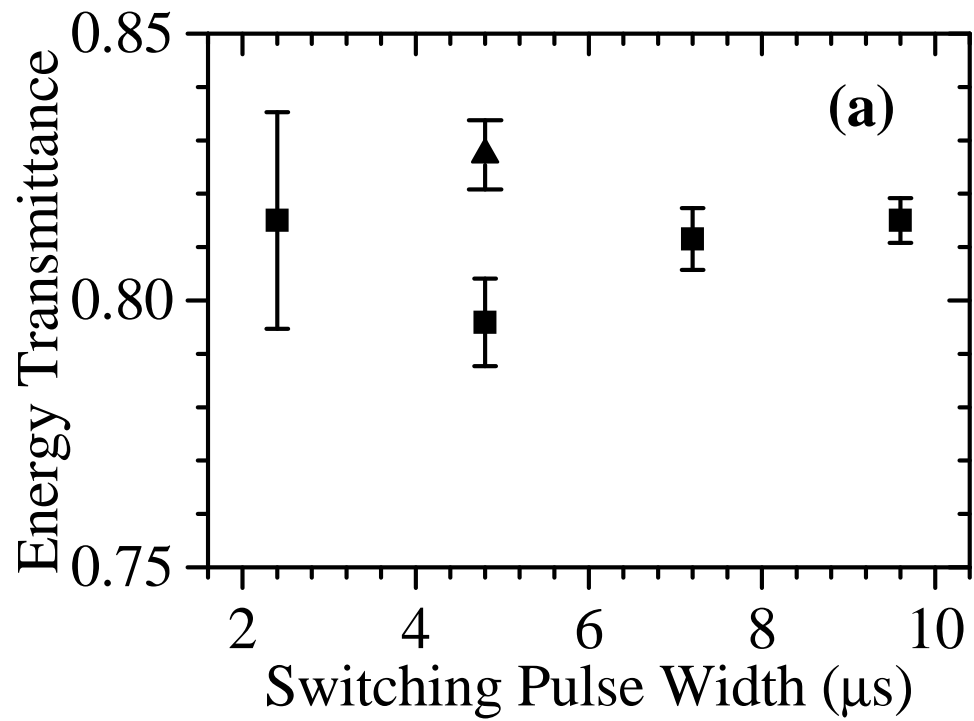
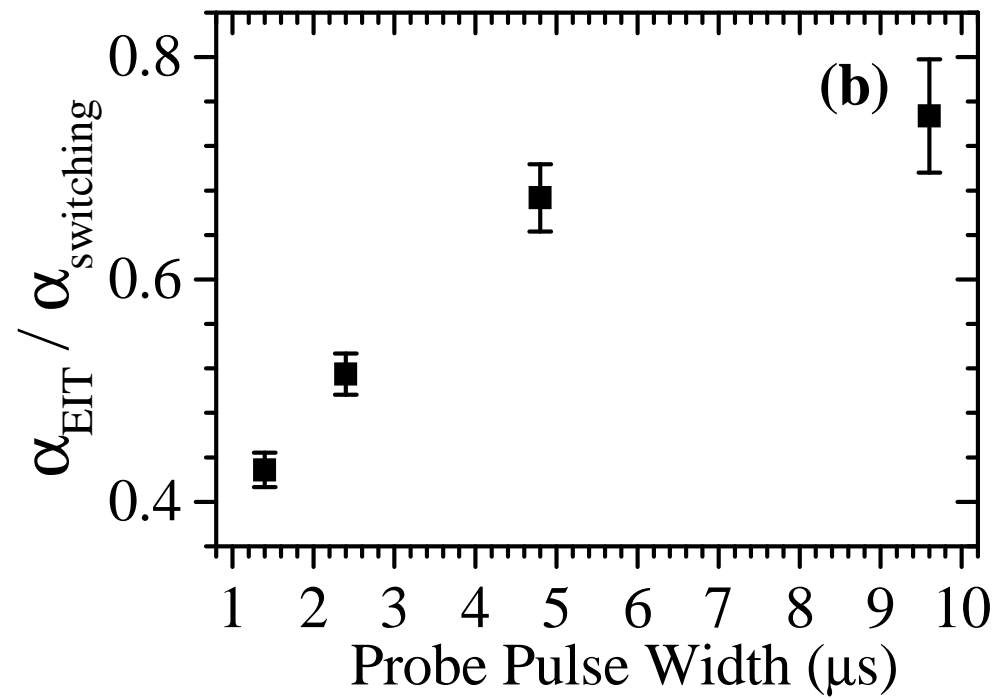
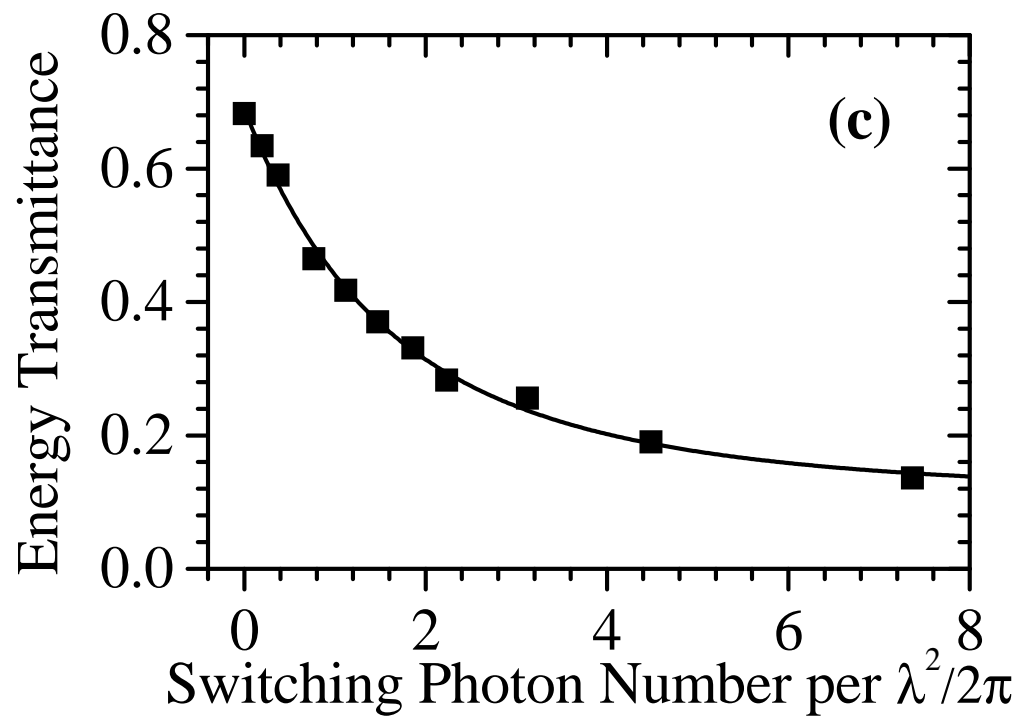
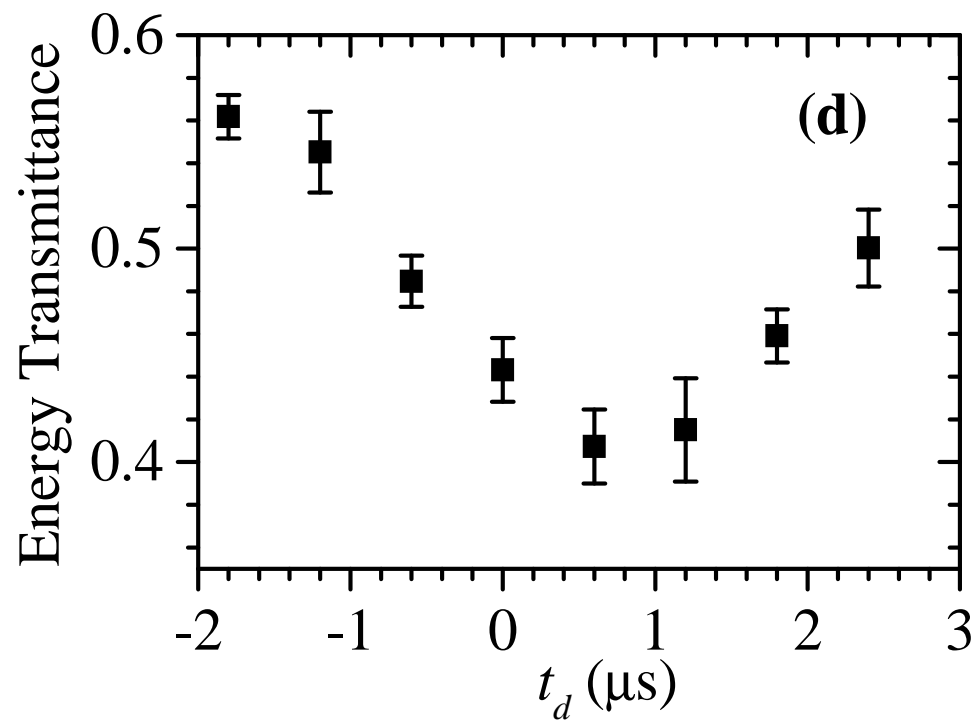
\end{document}